\def\abbrename{{\bfseries Abbreviations \runinend}}

\documentclass[global]{svjour}
\usepackage{graphics}
\usepackage{amssymb}
\journalname{Environmental Fluid Mechanics}
\begin{document}
\titlerunning{Subgrid-scale reactant flux model}
\authorrunning{Vinuesa et al.}
\title{Subgrid-scale modeling of reacting scalar fluxes in large-eddy simulations of atmospheric boundary layers}
\author{J.-F. Vinuesa\inst{1}\thanks{\emph{Corresponding address:} J.-F. Vinuesa, Saint Anthony Falls Laboratory, Mississippi river at 3rd Avenue SE, Minneapolis, MN 55414, USA, \email{vinue001@umn.edu}} \and F. Port{\'{e}}-Agel \inst{1,2} \and S. Basu \inst{1} \and R. Stoll\inst{1}
}                     
%
%
\institute{Saint Anthony Falls Laboratory and Department of Civil Engineering, University of Minnesota, Minneapolis, United States of America \and National Center for Earth-surface Dynamics}
\date{Submitted: June 17, 2005 } 

%
\maketitle
\begin{abstract}
In large-eddy simulations of atmospheric boundary layer turbulence, the lumped coefficient in the eddy-diffusion subgrid-scale 
(SGS) model is known to depend on scale for the case of inert scalars. This scale dependence is predominant near the surface.  
In this paper, a scale-dependent dynamic SGS model for the turbulent transport of reacting scalars is 
implemented in large-eddy simulations of a neutral boundary layer. Since the model coefficient is computed dynamically from 
the dynamics of the resolved scales, the simulations are free from any parameter tuning. 
A set of chemical cases representative of various turbulent reacting flow regimes is examined. The reactants
are involved in a first-order reaction and are injected in the atmospheric
boundary layer with a constant and uniform surface flux. 
Emphasis is placed on studying the combined effects of resolution and chemical 
regime on the performance of the SGS model. 
Simulations with the scale-dependent dynamic model yield the expected trends
of the coefficients as function of resolution, position in the flow and chemical regime, leading to
resolution-independent turbulent reactant fluxes.
\end{abstract}

\keywordname{Large-eddy simulation -- Subgrid-scale models -- Turbulent transport -- Atmospheric chemistry -- Atmospheric boundary layers}

\abbrename{ABL - Atmospheric boundary layer; LES - Large-eddy simulation; SGS - Subgrid scale; RANS - Reynolds Averaged Navier Stokes}

\section{Introduction}
\label{intro}
The atmospheric boundary layer (ABL) is the lower
part of the atmosphere, where turbulence plays an essential role on the transport of 
momentum and scalars (heat, water vapor and chemicals) between the
biosphere and the upper atmosphere. Quantifying the transport rate of chemicals
from the Earth's surface into the upper parts 
of the atmosphere is critical for a wide range of environmental applications, 
including air quality modeling and regional and global climate studies.
One important open question is how to represent the effects of boundary layer
physical processes, in particular turbulence, on chemical transport
and transformation of chemical compounds. A large number of reactive gases emitted 
by vegetation, such as biogenic
volatile organic compounds, have typical lifetimes that are 
of the same order of magnitude or smaller than the mixing time scales of
the largest ABL eddy motions \cite{kle97}. 
For such species (isoprene and monoterpenes for instance) the
interplay between turbulence and chemistry is of great importance. Accurate modeling 
of their chemical transformations in the ABL requires improved understanding on how 
turbulence and spatial distribution of the reactants affect chemistry.

Realistic numerical experiments on the turbulent transport and mixing of reactants 
require the use of Large-Eddy Simulation (LES). In the atmospheric boundary layer, the scales associated 
with turbulent motions range from the Kolmogorov dissipation scale (on the order of a millimeter) 
to the boundary layer depth (on the order of a kilometer). 
The largest eddies are responsible for the turbulent transport of the scalars and 
momentum whereas the smallest ones are mainly dissipative. LES consists of explicitly 
resolving all scales larger than the grid scale 
(on the order of tens of meters in the ABL), while the smallest (less energetic) 
scales are parameterized using a subgrid-scale (SGS) model. Since the pioneering paper of 
Schumann \cite{sch89} that showed for the first time that segregation of 
reactants due the inefficient mixing of the convective turbulence plays an
important role on moderating the reaction rates, several numerical studies on the effect of 
turbulence on chemistry in the ABL have been performed using LES 
\cite{gao93,syk94,gao94,ver97,mol98,pet99b,pet99c,pet00,kro00,ver00,vil00,pat01,vin03,jon04,vin04}. 
In particular, Gao {\it et al.} \cite{gao93} and Patton {\it et al.} \cite{pat01} focused
their study on the lowest part of the ABL and on the interplay between chemistry and turbulence in and above
canopies.

An important challenge in large-eddy simulations of the atmospheric boundary
layer is the specification of the SGS model coefficients
and, in particular, how to account for external or internal parameters such as position in the flow,
grid/filter scale, atmospheric stability and chemical regime.
Usually in LES of atmospheric turbulent reacting flows, the subgrid reactant flux is
modelled as the product of a local gradient of the concentrations 
from the resolved scales and of eddy-diffusity coefficients. These coefficients are either simply prescribed (in the
case of Smagorinsky models)
or estimated
by solving an equation for the turbulent kinetic energy assuming that subgrid scale turbulence is equally efficient
at transporting scalars as it is for heat \cite{moe84}.
In this paper, we explicitly calculate the model coefficients by using a dynamic procedure and without making
any assumption of the ability of subgrid turbulence to transport scalars. Thus the model coefficients are
specifically calculated for each reacting scalar. 

When studying the effects of the unresolved scales on the chemical transformations, e.g. related to the subgrid scale mixing
between the reactants
due to subgrid scale turbulence, Vinuesa and Port{\'e}-Agel \cite{vin05} suggest
that these effects can be correlated with the chemical regime. Since the SGS reactant flux includes
a chemical contribution in its governing equation, chemical regime dependency of the subgrid scale transport can be also
expected. By using a set of chemical setups representative of several chemical regimes, 
one can study the relevance of this chemical dependency of the reactant subgrid fluxes.

Recently, Port{\'e}-Agel \cite{por04}
showed that the lumped coefficient used in the eddy-diffusion models for inert
scalars are strongly scale dependent, in particular close to the surface. In order to account for that 
scale dependence of the coefficient, he developed a scale-dependent 
dynamic procedure to compute the value of the eddy-diffusion coefficient using 
information contained in the resolved scales. In this paper, we use this tuning-free 
scale-dependent dynamic model for simulations of reacting scalars.

The structure of this paper is as follows. In
section \ref{sec2}, we present a scale-dependent dynamic model for the turbulent transport
of reactants in the ABL. The model is used for simulations
of a neutral atmospheric boundary layer. The characteristics of the simulations are presented in section \ref{sec3}. 
A set of simulations is performed in order to study the combined effects of chemical regime and resolution. 
The relative importance of subgrid-scale turbulence 
on the chemistry is classified using a subgrid-scale Damk\"{o}hler number. 
By focusing our analysis on a first order decaying scalar, we are able to isolate the effects of 
the unresolved scales on the turbulent transport of the reactants. 
Emphasis is placed on the scale-independence of the simulated reactant fluxes and on 
the combined effects of chemical regime and resolution on 
the dynamically computed model coefficients and scale-dependence parameters. 
Finally, conclusions are drawn in section \ref{sec2}.

\section{The scale-dependent dynamic model}
\label{sec2}
The filtered governing equation for the concentration of a reacting scalar $A$
that is solved in LES reads: 

\begin{equation}
\frac{\partial {\widetilde{c_A}}}{\partial{t}}+\widetilde{u}_{i}\frac{
\partial {\widetilde{c_A}}}{\partial{x_{i}}}=-\frac{\partial{\widetilde{u^{\prime\prime}_{i}c_A^{\prime\prime}}}}{
\partial {x_{i}}} +\widetilde{R}_{ch}
\label{eq3.0}
\end{equation}

\noindent where $\widetilde{c_A}$ and $c_A^{\prime\prime}$ are the spatially filtered (at the filter scale $\Delta $)
and the subgrid concentrations of the reactant, respectively. $\widetilde{u^{\prime\prime}_{i}c_A^{\prime\prime}}$ 
is the subgrid-scale flux and $\widetilde{R}_{ch}$ is the chemical
term. For a reactant $A$ involved in the first-order reaction

\begin{eqnarray}
A~\stackrel{j}{\rightarrow} ~ Products,  \label{cond3}
\end{eqnarray}

\noindent the chemical term reads 

\begin{eqnarray}
\widetilde{R}_{ch}&=&-j\widetilde{c_A}. \label{eq3.01}
\end{eqnarray}

The effect of the
unresolved scales on the evolution of the filtered scalar concentration
appears through the subgrid scale reactant flux

\begin{equation}
\widetilde{u^{\prime\prime}_{i}c_A^{\prime\prime}}=\widetilde{u_{i}c_A}-\widetilde{u}_{i}\widetilde{c_A}.  \label{eq3.1}
\end{equation}

It is important to note that in the case of a second-order reaction (not
considered here), the unresolved scales can affect the reactant
concentrations not only through the subgrid-scale fluxes but also through
the subgrid-scale covariances (see e.g. \cite{mee00,vin05}).

The subgrid reactant flux is usually
modeled as a function of a local gradient of the resolved concentrations 
using an eddy-diffusity model. The subgrid-scale Schmidt
number in that model is either simply prescribed as a constant value or
estimated assuming that subgrid scale turbulence is equally efficient at
transporting reacting scalars as it is for heat.

The eddy-diffusion subgrid model is widely used in LES of the atmospheric boundary
layer. A common formulation of this model for the subgrid scale flux of a scalar $\psi$ is 

\begin{equation}
\widetilde{u^{\prime\prime}_{i}c_\psi^{\prime\prime}}=-\Delta^{2}\left[Sc_{sgs,\psi}^{-1}C_{S}^{2}(\Delta )\right] 
\left| \widetilde{S}\right| \frac{\partial \widetilde{c_\psi}}{\partial
x_{i}}.  \label{eq3.2}
\end{equation}

\noindent where $|\widetilde{S}|=(2\widetilde{S}_{ij}\widetilde{S}_{ij})^{1/2}$ is the
resolved strain-rate magnitude, $\widetilde{S}_{ij}$ is the resolved
strain rate tensor, $\Delta$ is the filter scale and $\widetilde{c_\psi}$ is the resolved (spatially filtered) 
scalar concentration. 
$C_{S}$ and $Sc_{sgs,\psi}$ denote the Smagorinsky coefficient in the eddy-viscosity model
and the subgrid-scale Schmidt number for the scalar $\psi$, respectively. 

Usually the subgrid-scale Schmidt number $Sc_{sgs,\psi}$ and, as a
consequence, the lumped coefficient $Sc_{sgs,\psi }^{-1}C_{S}^{2}$ are only
determined (calculated or prescribed) for inert scalars, and those values are
used for the calculation of the subgrid scale reactant fluxes using Equation~(\ref{eq3.2}).
It is, therefore, assumed that chemistry does not affect the
subgrid scale fluxes and, consequently, $Sc_{sgs,\psi }=Sc_{sgs}$, where 
$Sc_{sgs}$ is the subgrid-scale Schmidt number for an inert scalar. $Sc_{sgs}$
is often taken as a constant value ($Sc_{sgs}\approx 1/3$), which is
well-established for isotropic turbulence \cite{lil67,mas90}. However,
near the surface of ABL flows, the reacting scalar field at the smallest
resolved scales can become anisotropic due to the combined effects of
proximity to the surface, resolution (filter/grid scale), atmospheric
stability and/or chemical regime.  In the following, we propose an
alternative procedure to calculate the value of the lumped eddy-diffusion
model coefficient that is able to adjust to the anisotropy of the reacting
scalar field associated with the aforementioned factors. In particular, the
value of the lumped model parameter $Sc_{sgs,\psi }^{-1}C_{S}^{2}$ is
determined based on the dynamics of the smallest resolved scales using the
scale-dependent dynamic model introduced for passive scalars by Port\'{e}-Agel \cite{por04}.
This tuning-free procedure is summarized below.

For the reactant subgrid scale flux $\widetilde{u^{\prime\prime}_{i}c_A^{\prime\prime}}$ (denoted $q_{A,i}$ hereafter
to enhance readability), the dynamic procedure is based on the Germano identity for scalars \cite{ger92,lil92}

\begin{equation}
K_{A,i}=Q_{A,i}-\overline{q}_{A,i}=\overline{\widetilde{u}_{i}\widetilde{c_A}}-
\overline{\widetilde{u}}_{i}\overline{\widetilde{c_A}},  \label{eq3.3}
\end{equation}

where $Q_{A,i}=\overline{\widetilde{u_{i}c_A}}-\overline{\widetilde{u}}_{i}
\overline{\widetilde{c_A}}$ is the subgrid scale flux at a test-filter scale
(typically taken as $\overline{\Delta }=2\Delta $) and $K_{A,i}$ is a resolved reactant flux
that can be evaluated based on the resolved scales. Applying the
eddy-diffusion model, $Q_{A,i}$ is determined by 

\begin{equation}
Q_{A,i}=-\left[ Sc_{sgs,A}^{-1}C_{S}^{2}\left( \overline{\Delta }\right) 
\right] \overline{\Delta}^{2}\left| \overline{\widetilde{S}}\right| \frac{
\partial \overline{\widetilde{c_A}}}{\partial x_{i}}.  \label{eq3.4}
\end{equation}

Substitution of Equations. (\ref{eq3.2}) and (\ref{eq3.4}) into (\ref{eq3.3}) leads to
the system 

\begin{equation}
K_{A,i}=Sc_{sgs,A}^{-1}C_{S}^{2}X_{A,i},  \label{eq3.5}
\end{equation}

where, for $\overline{\Delta }=2\Delta$,

\begin{equation}
X_{A,i}=\Delta ^{2}\left( \overline{\left| \widetilde{S}\right| \frac{\partial 
\widetilde{c_A}}{\partial x_{i}}}-4\frac{Sc_{sgs,A}^{-1}C_{S}^{2}\left(
2\Delta \right) }{Sc_{sgs,A}^{-1}C_{S}^{2}\left( \Delta \right) }\left| 
\overline{\widetilde{S}}\right| \frac{\partial \overline{\widetilde{c_A}}
}{\partial x_{i}}\right) .  \label{eq3.6}
\end{equation}

Note that the traditional dynamic model \cite{por04} assumes scale
invariance of the model coefficient at the filter and test filter scales,
i.e., 

\begin{equation}
Sc_{sgs,A}^{-1}C_{S}^{2}\left( \Delta \right)
=Sc_{sgs,A}^{-1}C_{S}^{2}\left( 2\Delta\right)
=Sc_{sgs,A}^{-1}C_{S}^{2}.  \label{eq3.7}
\end{equation}

Minizing the error associated with the use of the eddy-diffusion model in
Equation~(\ref{eq3.3}) over all three vector components results in 

\begin{equation}
Sc_{sgs,A}^{-1}C_{S}^{2}\left( \Delta \right) =\frac{\left\langle K_{A,i}X_{A,i}\right\rangle }
{\left\langle X_{A,i}X_{A,i}\right\rangle },  \label{eq3.8}
\end{equation}

\noindent where $\left\langle{ }\right\rangle$ denotes an averaging operator 
 (typically over horizontal planes) and
$X_{A,i}$ contains the scale-dependent parameter $\beta_{A}=\frac{Sc_{sgs,A}^{-1}C_{S}^{2}\left(
2\Delta \right) }{Sc_{sgs,A}^{-1}C_{S}^{2}\left( \Delta \right) }$. Please note that $\beta_{A}=1$ in
the case of the traditional dynamic model.
A dynamic procedure can be developed to compute the
value for $\beta _{A}$ using a second test-filtering operation at scale $\widehat{\Delta }>\overline{\Delta }$.
For the sake of simplicity, we take a second test-filter scale as $\widehat{\Delta }=4\Delta $, 
and denote variables filtered at scale $4\Delta $ by a caret $\widehat{ }$. This
results in a second equation for $Sc_{sgs,A}^{-1}C_{S}^{2}\left( \Delta \right)$ that reads 

\begin{equation}
Sc_{sgs,A}^{-1}C_{S}^{2}\left( \Delta \right) =\frac{\left\langle
K_{A,i}^{\prime }X_{A,i}^{\prime }\right\rangle }{\left\langle X_{A,i}^{\prime
}X_{A,i}^{\prime }\right\rangle },  \label{eq3.12}
\end{equation}

\noindent where 

\begin{equation}
K_{A,i}^{\prime }=\widehat{\widetilde{u}_{i}\widetilde{c_A}}-\widehat{
\widetilde{u}}_{i}\widehat{\widetilde{c_A}},\label{eq3.10}
\end{equation}

\noindent and 

\begin{equation}
X_{A,i}^{\prime }=\Delta ^{2}\left( \widehat{\left| \widetilde{S}\right| \frac{
\partial \widetilde{c_A}}{\partial x_{i}}}-4^{2}\frac{
Sc_{sgs,A}^{-1}C_{S}^{2}\left( 4\Delta \right) }{Sc_{sgs,A}^{-1}C_{S}^{2}
\left( \Delta \right) }\left| \widehat{\widetilde{S}}\right| \frac{\partial 
\widehat{\widetilde{c_A}}}{\partial x_{i}}\right) .  \label{eq3.11}
\end{equation}

By combining Equations.~\ref{eq3.8}~and~\ref{eq3.12},
one obtains the following equation from which $\beta_{A}$ can be computed (more details on this
procedure are given in Port\'{e}-Agel \cite{por04}), 

\begin{equation}
\left\langle K_{A,i}X_{A,i}\right\rangle \left\langle X_{A,i}^{\prime
}X_{A,i}^{\prime }\right\rangle -\left\langle K_{A,i}^{\prime }X_{A,i}^{\prime
}\right\rangle \left\langle X_{A,i}X_{A,i}\right\rangle =0.
\label{eq3.13}
\end{equation}

The value of $\beta_{A}$ is extracted from Equation~\ref{eq3.13} assuming that

\begin{equation}
\frac{Sc_{sgs,A}^{-1}C_{S}^{2}
\left( 2\Delta \right)} {Sc_{sgs,A}^{-1}C_{S}^{2}
\left( \Delta \right)} =\frac{Sc_{sgs,A}^{-1}C_{S}^{2}
\left( 4\Delta \right)}{Sc_{sgs,A}^{-1}C_{S}^{2}
\left( 2\Delta \right)} 
\end{equation}

\noindent which implies that $Sc_{sgs,A}^{-1}C_{S}^{2}
\left( 4\Delta \right)/Sc_{sgs,A}^{-1}C_{S}^{2}
\left( \Delta \right)  = \beta_{A}^{2}$. 

Then $\beta_{A}$ is used in Equation~\ref{eq3.8} to obtain
the lumped coefficient $Sc_{sgs,A}^{-1}C_{S}^{2}$. 

\section{Reacting atmospheric flow simulations}
\label{sec3}
\subsection{Turbulent reacting flow classification}
\label{sec3.1}

Turbulent reacting flows can be classified by using two dimensionless numbers 
\cite{sch89,vil03a}. The first
one, i.e. the turbulent integral Damk\"{o}hler number, $Da_{t}$, refers to the influence of the largest atmospheric
boundary layer eddies on reacting scalars. In the context of Reynolds Averaged Navier Stokes (RANS) modeling,
it gives a measure of
the relative importance of the dynamical and chemical contributions to the
reacting scalar mean governing equation. The second number,
i.e. the Kolmogorov Damk\"{o}hler number, $Da_{k}$, quantifies the
relative magnitude of the characteristic time scales of the smallest ABL
eddies with respect to the chemical reaction time scale. These numbers are defined as

\begin{equation} \label{eq2.2}
Da_{t}=\frac{\tau_t}{\tau_c}, \end{equation}

\begin{equation} \label{eq2.3}
Da_{k}=\frac{\tau_k}{\tau_c}, \end{equation}

\noindent where $\tau_t$ is the integral turbulent time scale, $\tau_k$ is the Kolmogorov time scale and $\tau_c$
is the chemical time scale. 

Using these Damk\"{o}hler numbers, the following classification of chemical regime 
in turbulent reacting flows can be made:
\begin{itemize}
    \item $Da_{t}<1$, slow chemical regime: the reactant is uniformly mixed in the 
    boundary layer and its governing equation is dominated by the dynamical terms.
     \item $Da_{k}<1<Da_{t}$, moderate chemical regime: the largest scales affect the turbulent mixing
     whereas the chemistry is not limited by the small-scale turbulence.
     \item $Da_{k}>1$, fast chemical regime: turbulence controls the chemistry at all scales.
\end{itemize}    

In the context of large-eddy simulation, it is also very important
to assess the effects of the subgrid scales on chemistry. In order to
achieve that, it is convenient to introduce the so-called subgrid 
Damk\"{o}hler number or $Da_{sgs}$ \cite{kro00} 

\begin{equation}\label{eq2.40}
Da_{sgs}=\frac{\tau _{sgs}}{\tau _{c}},
\end{equation}

$Da_{sgs}$ is a measure of the relative magnitude of 
the time scale associated with the smallest eddies
resolved in LES with respect to the chemical time
scale. That time scale can be computed as

\begin{equation} \label{eq2.4}
\tau_{sgs}=\left(\frac{\nu_{sgs}}{\epsilon_{sgs}}\right)^{1/2}, \end{equation}

\noindent where $\nu_{sgs}$ and $\epsilon_{sgs}$ are the subgrid-scale
eddy viscosity and the subgrid-scale dissipation rate of the turbulent
kinetic energy, respectively. The subgrid-scale dissipation rate is defined
as the rate of transfer of kinetic energy between the resolved and the
subgrid scales \cite{pop00,men00}. 
By using $Da_{sgs}$, turbulent
reacting flows are classified with respect to the effect of the unresolved scales on the chemistry.
For $Da_{sgs}<<1$, the flow is well-mixed at subgrid scales and, consequently, the unresolved
scales do not influence the behavior of reacting scalars. For $Da_{sgs} \sim O(1)$ and $>1$, all
the scales of turbulence, i.e. resolved and subgrid, affect the chemistry. These
effects on the transport and mixing of reactants lead to a heterogeneous distribution of the reactants
at subgrid scales.

\subsection{Numerical set-up}
\label{sec3.2}

The code used is a modified version of 3-dimensional LES code described by Albertson and Parlange \cite{alb99},Port\'{e}-Agel {\it et al.} \cite{por00}, and Port\'{e}-Agel \cite{por04}. Briefly, the code uses a mixed 
pseudospectral finite-difference method and the subgrid-scale stresses and momentum and temperature fluxes are parameterized 
with scale-dependent dynamic models. Test filtering for dynamic models 
is done using two-dimensional spectral cutoff filters. A chemical solver has been implemented in the LES code
\cite{vin05}.
We simulate a neutral atmospheric boundary layer driven by a pressure gradient forcing
and a surface roughness length for momentum of $0.1m$ is prescribed.
The computational domain is of size ($L_x$, $L_y$, $L_z$) and it correspond to 
a grid prescribed with $N$ x $N$ x $N$ points. The height of the
domain $L_z$ is equal to $1000$ meters and $L_x = L_y = 2 \pi L_z$. 
Several resolutions are used with $N$ equal to $32$, $64$, and $128$.

In order to restrict our study to the performance of the scale dependent subgrid-scale model
for the reactant flux, the studied reacting flow consists of the first-order reaction (\ref{cond3}).

We define three chemical setups based on this first-order irreversible reaction. 
The reactant $A$, so-called bottom-up reactant, is uniformly emitted at the 
surface with a flux of $0.1~ppb~m~s^{-1}$ and no initial
concentrations in the boundary layer. The reaction rate $j$ is set to zero,
$4.5$~x~$10^{-4}~s^{-1}$, and $9$~x~$10^{-3}~s^{-1}$. The corresponding chemical
set-up will be refer to as inert ($I$), slow ($S$) and fast ($F$) chemical cases. 
Notice that, in our simulations, $\tau_t=h/u_*$ and $\tau_c=j^{-1}$
lead to turbulent Damk\"{o}hler numbers $Da_{t}$ equals to $0$, $1$ and $20$, respectively.
 The simulations cover a $1.5$ hours period
and the statistics presented here are obtained averaging the
results over the last hour of simulation. 

\subsection{Results}
\label{sec3.3}

\begin{figure}
\resizebox{0.75\textwidth}{!}{%
  \includegraphics{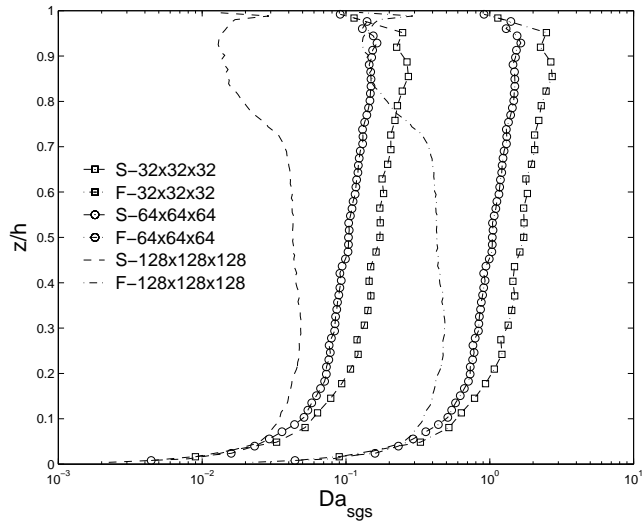}
}
\caption{Vertical profiles
of the subgrid scale Damk\"{o}hler numbers. These numbers have been obtained dynamically and
averaged over the last hour of simulation. The $S$ and $F$ chemical cases
are represented by dashed and dash-dotted lines, respectively, 
and the $32^{3}$, $64^{3}$ and $128^{3}$ grid-points simulations are plotted 
with squares, circles and no symbol, respectively.} \label{fig:fig1}
\end{figure}

In Fig.~\ref{fig:fig1}, the vertical profiles of the subgrid scale Damk\"{o}hler number $Da_{sgs}$ are shown.
For the slow ($S$) chemical regime, the chemical time scale ($\tau _{c}$) is
much larger than the characteristic subgrid-scale time scale ($\tau _{sgs}$), 
which determines a relatively small value of $Da_{sgs}$ for all
resolutions. On the contrary, all the fast ($F$) chemical cases have $Da_{sgs}\sim O(1)$ 
and, as a result, the transport of the reacting scalar
is affected by the subgrid scale chemistry. The value of $Da_{sgs}$ and,
consequently, the effect of the subgrid-scale chemistry on the scalar
fluxes, increases with decreasing resolution due to the fact that the
characteristic time scale of the subgrid fluxes is larger for coarser grids.

\begin{figure}
\resizebox{0.75\textwidth}{!}{%
  \includegraphics{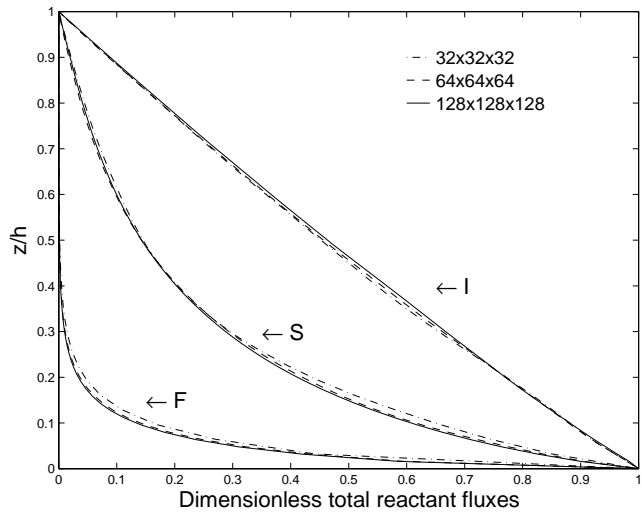}
}
\caption{Vertical profiles
of the total scalar fluxes averaged over the last hour of simulation. Results are made dimensionless
by using the surface fluxes.} \label{fig:fig2}
\end{figure}

Fig.~\ref{fig:fig2} shows the vertical profiles of the time-averaged total
scalar fluxes obtained as the sum of the resolved and the
subgrid-scale fluxes. 
Within the boundary layer, the profiles
of inert scalar fluxes have a linear shape. As noticed by Port\'{e}-Agel \cite{por04},
since a constant mean surface flux is imposed and no viscous effects are considered,
the averaged total turbulent flux decreases linearly with height. The reactive scalar flux profiles,
however, deviate from this linear behavior. In fact, these deviations become
more significant when the turbulent Damk\"{o}hler number increases; they are
larger for the $F$ experiment than for the $S$ one. The deviations increase with 
the reaction rate and, thus, with the turbulent Damk\"{o}hler number due to the increase 
in the chemical contribution to the flux. This was shown also
by Gao and Wesely \cite{gao94}, Sykes {\it et al.} \cite{syk94}, and Vinuesa and Vil\`{a}-Guerau de Arellano \cite{vin03} for
convective boundary layers. 

\begin{figure}
\resizebox{0.75\textwidth}{!}{%
  \includegraphics{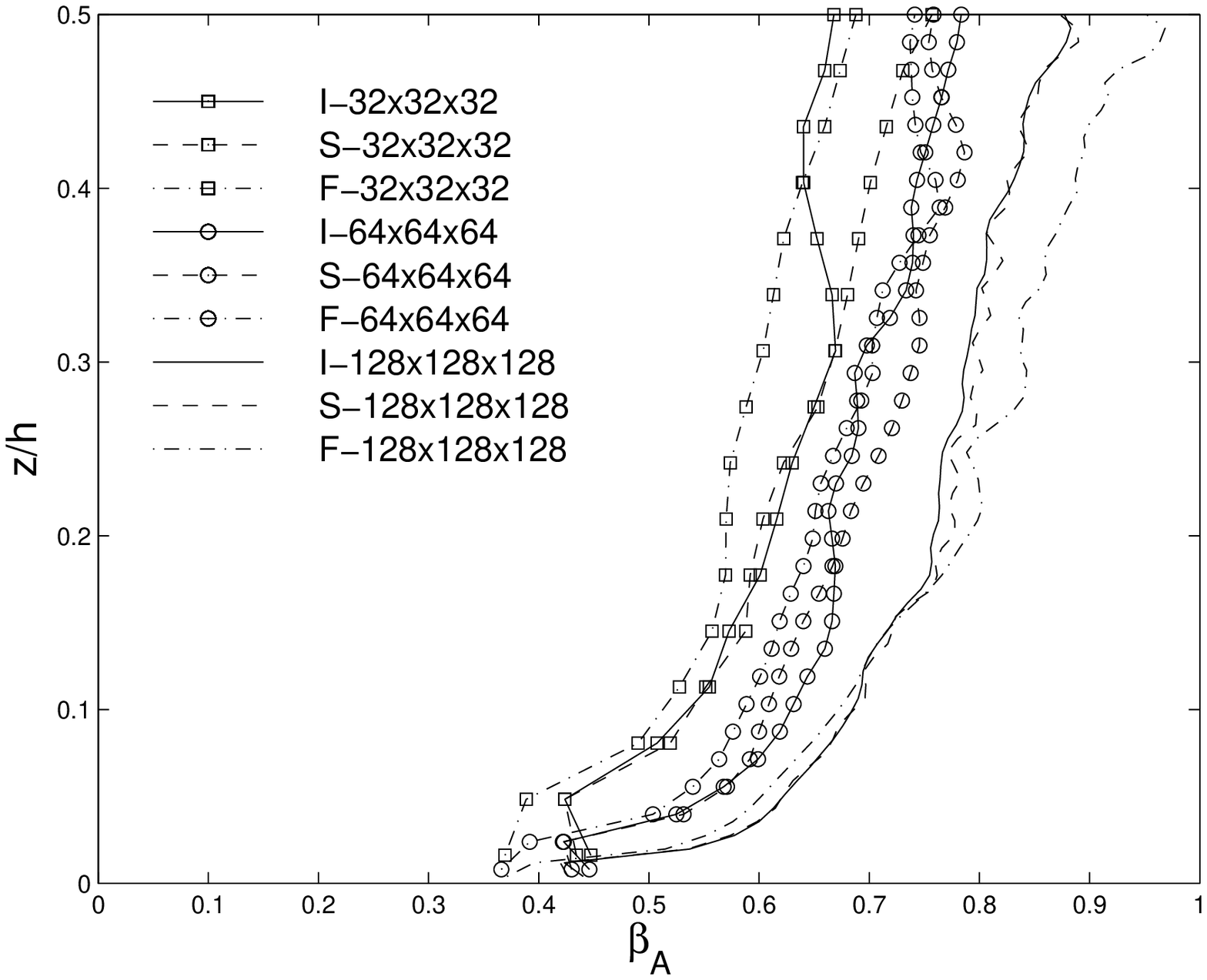}
}
\resizebox{0.75\textwidth}{!}{%
  \includegraphics{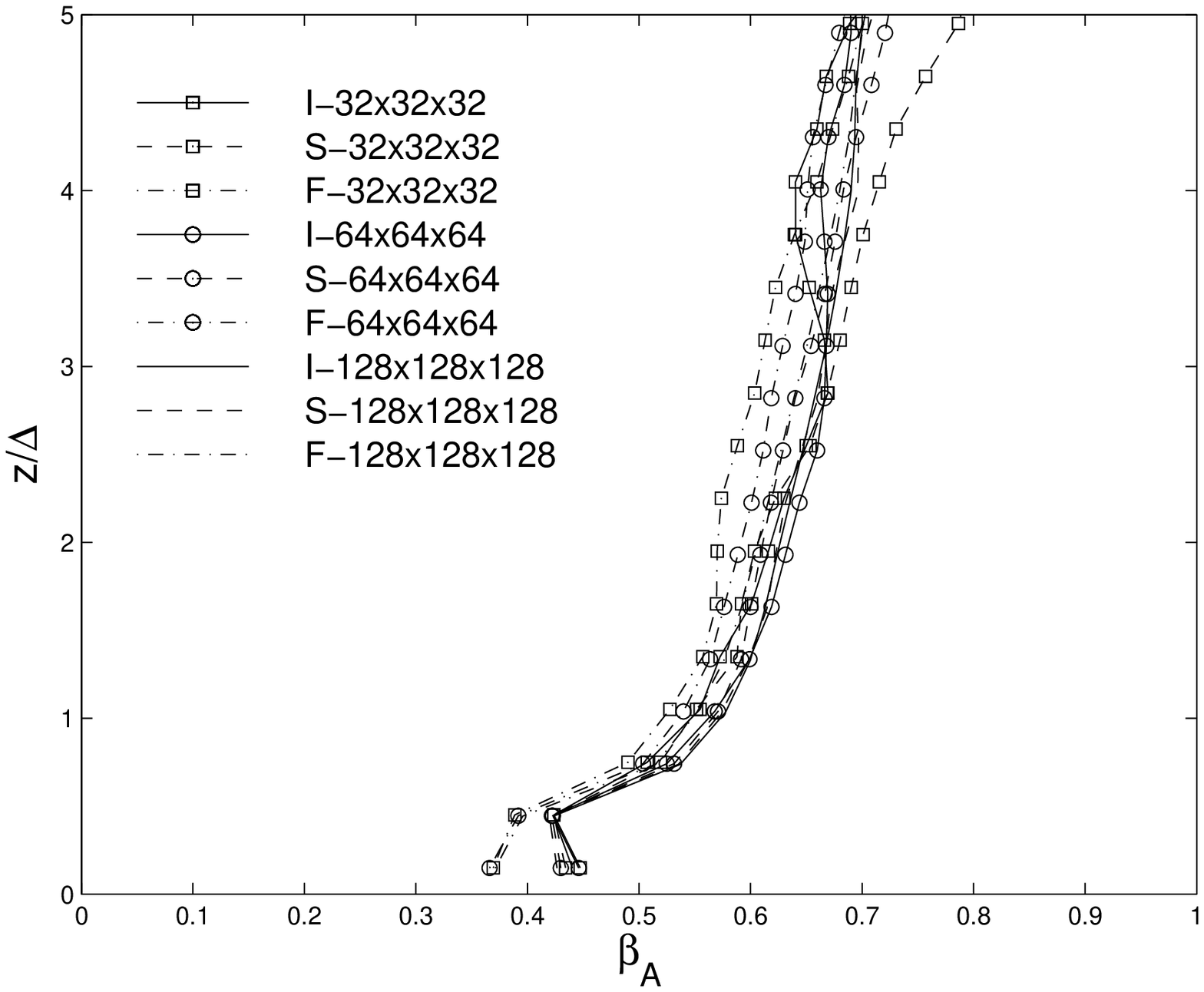}
}
\caption{Vertical profiles of $\beta_{A}$ as a function of the
height made dimensionless using the boundary layer depth $h$ (upper figure) and 
the filter size $\Delta$ (lower figure). The results have been averaged over the last hour of simulation.} \label{fig:fig3}
\end{figure}

As mentioned earlier, the magnitude of the subgrid-scale fluxes are affected
by both the resolution and the chemical regime. The fact that the total
turbulent fluxes (resolved plus subgrid contributions) for each chemical
regime show almost identical results for the three resolutions under
consideration highlights the robustness of the scale-dependent dynamic
model. This subgrid model is able to adjust the magnitude of the
subgrid-scale fluxes to yield the same total turbulent fluxes. It is
important to note that the resolution independence of the results is one of the
most challenging tests for any subgrid-scale model, especially when
the subgrid-scale effects are also affected by the chemical regime.

The time-averaged values of the scale-dependence
parameter $\beta_{A}$ are shown in Fig.~\ref{fig:fig3} as a
function of the normalized heights $z/h$ and $z/\Delta $.
$\beta _{A}$ is smaller near the
surface, indicating a strong scale dependence in that region, and it
increases to larger values far from the surface. Note, however, that the
scale-dependence coefficients obtained far from the surface are still
slighly smaller than the value of 1, indicating a slight scale dependence
and flow anisotropy. This behavior contrasts with the value of 1 obtained
for the scale-dependent parameter for the momentum fluxes as reported by 
Port\'{e}-Agel \cite{por04}, and it could be attributed to the more anisotropic
behavior of scalars (compared with the velocity field) in turbulent flows.
From Fig. \ref{fig:fig3}, it is also clear that $\beta _{A}$ depends on both
resolution (scale $\Delta $) and chemical regime (characterized by the
subgrid Damk\"{o}hler number). The scale dependence of $\beta _{A}$ was
already demonstrated by Port\'{e}-Agel \cite{por04} for the case of an inert
scalar, by showing that the values of $\beta _{A}$ computed at different
resolutions collapse when plotted against $z/\Delta $. The fact that in our
simulations the collapse of the curves is not perfect is due to the
additional effect of chemistry on the subgrid-scale fluxes. This effect
is further explored below.

\begin{figure}
\center
\resizebox{0.75\textwidth}{!}{%
  \includegraphics{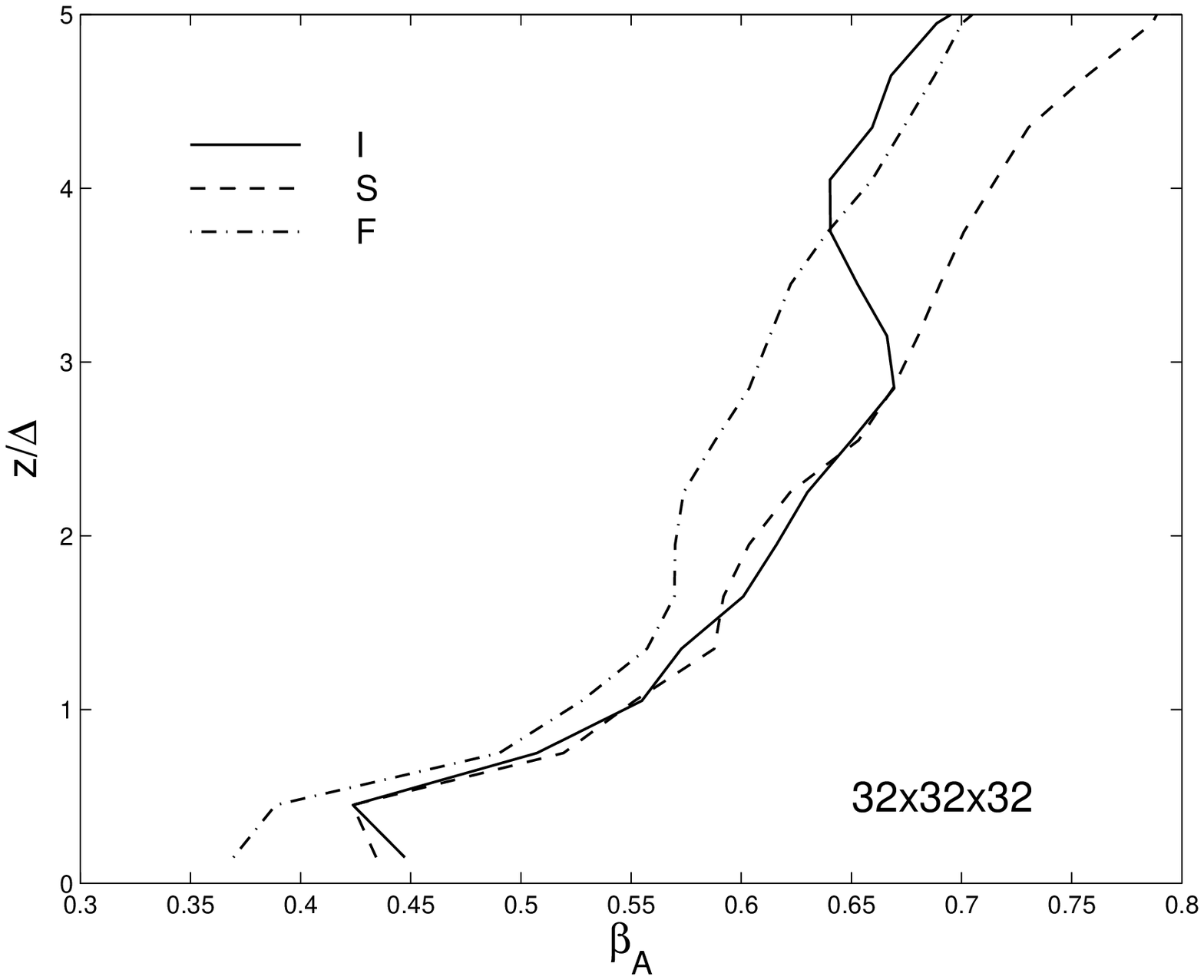}\includegraphics{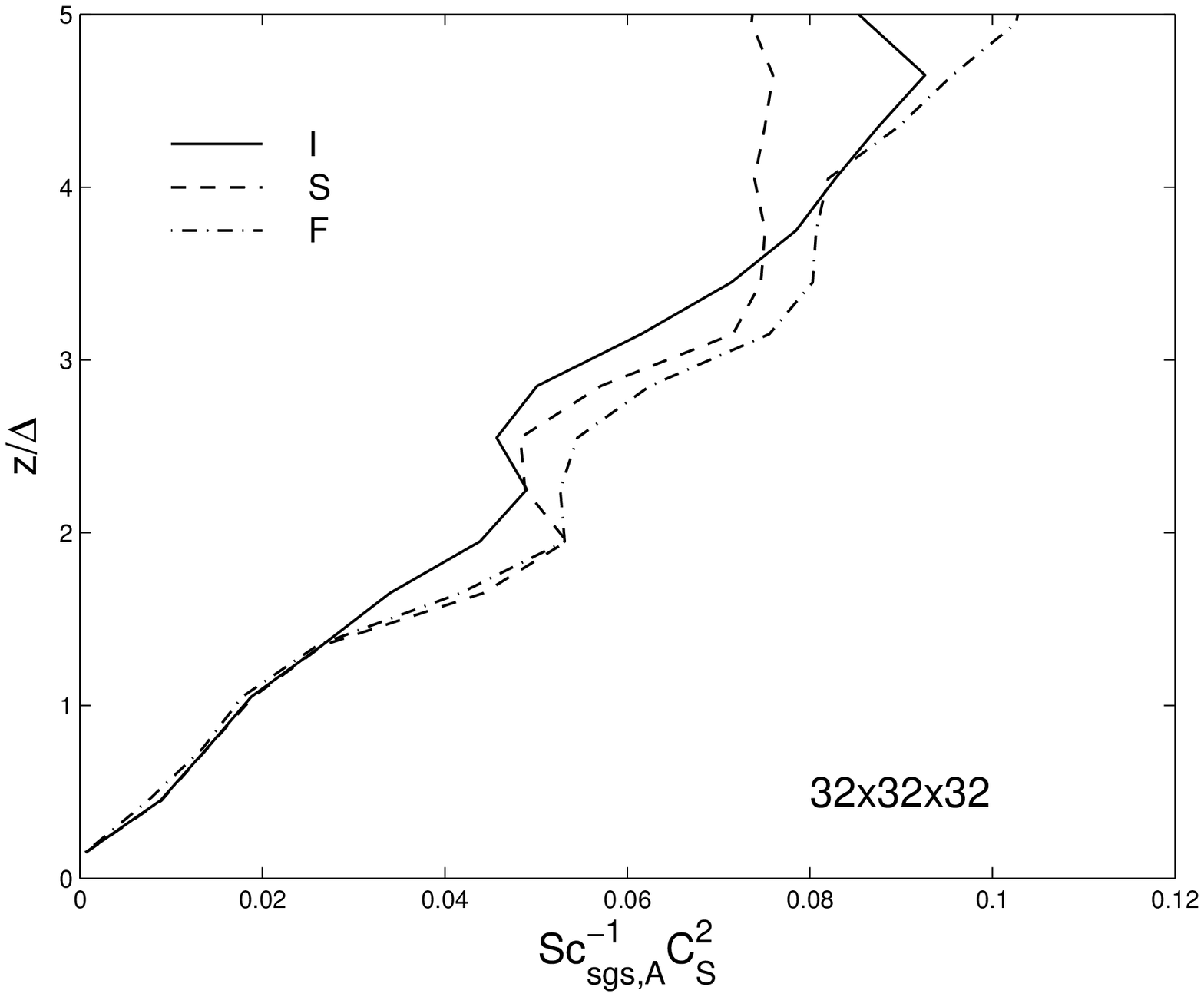}
}
\resizebox{0.75\textwidth}{!}{%
  \includegraphics{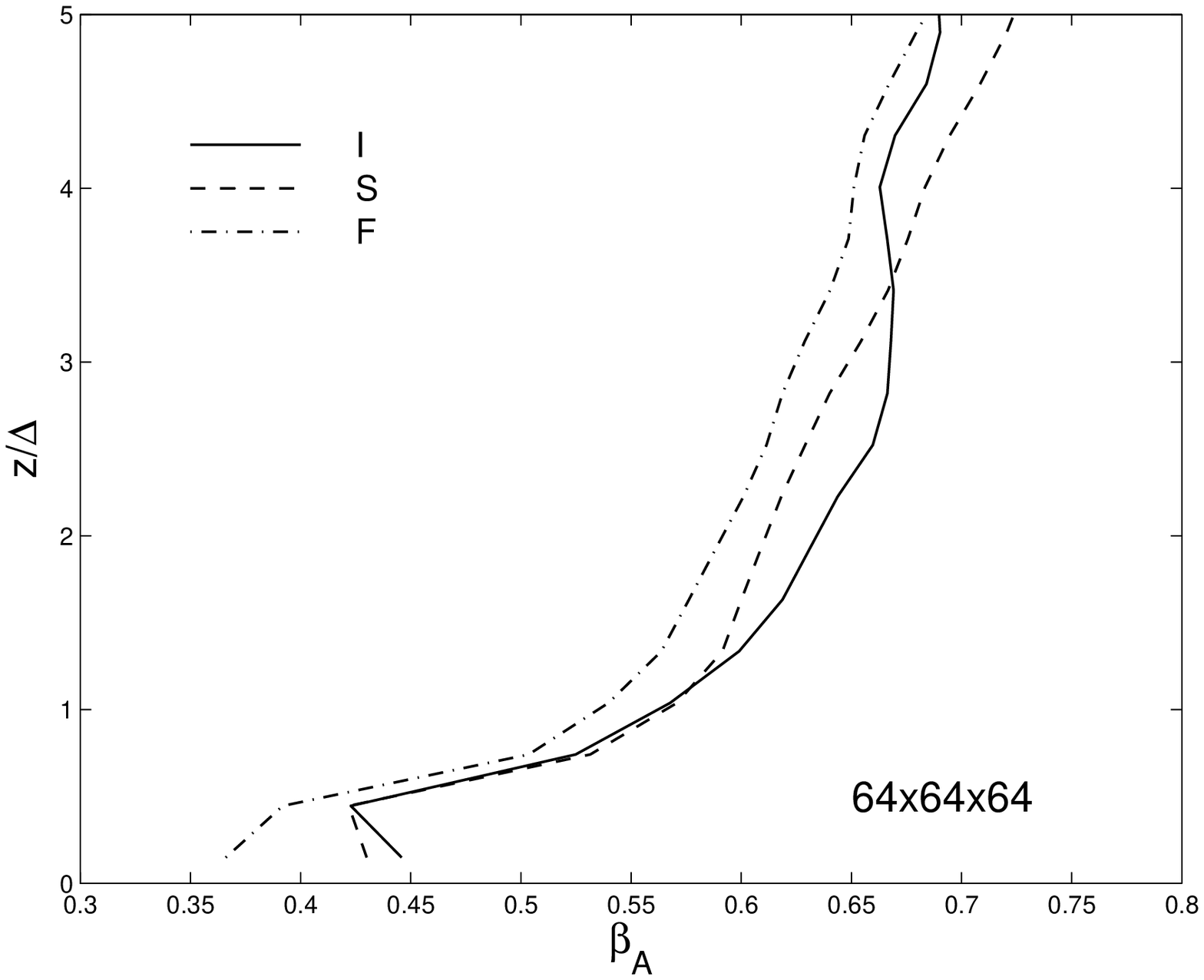}\includegraphics{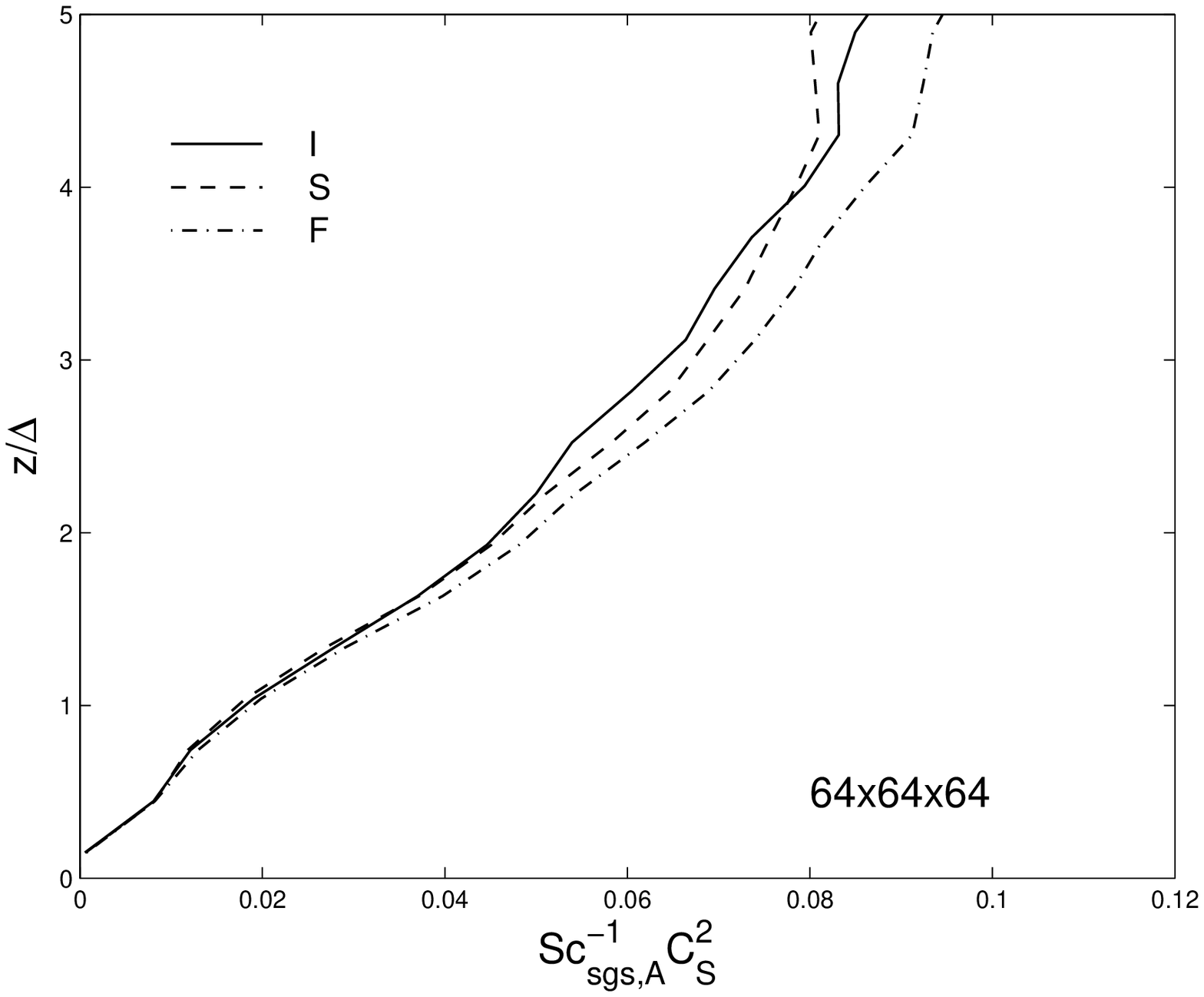}
}
\resizebox{0.75\textwidth}{!}{%
  \includegraphics{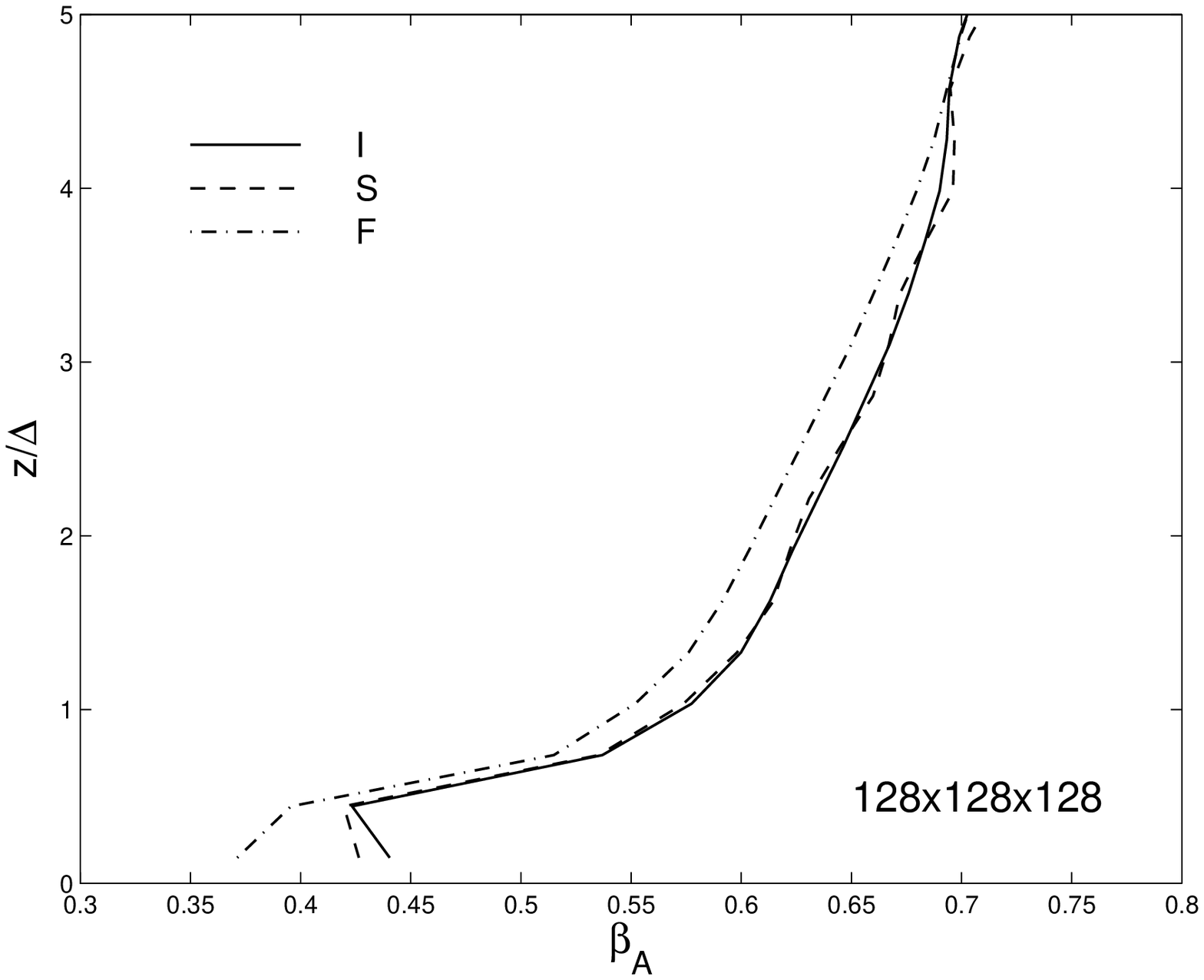}\includegraphics{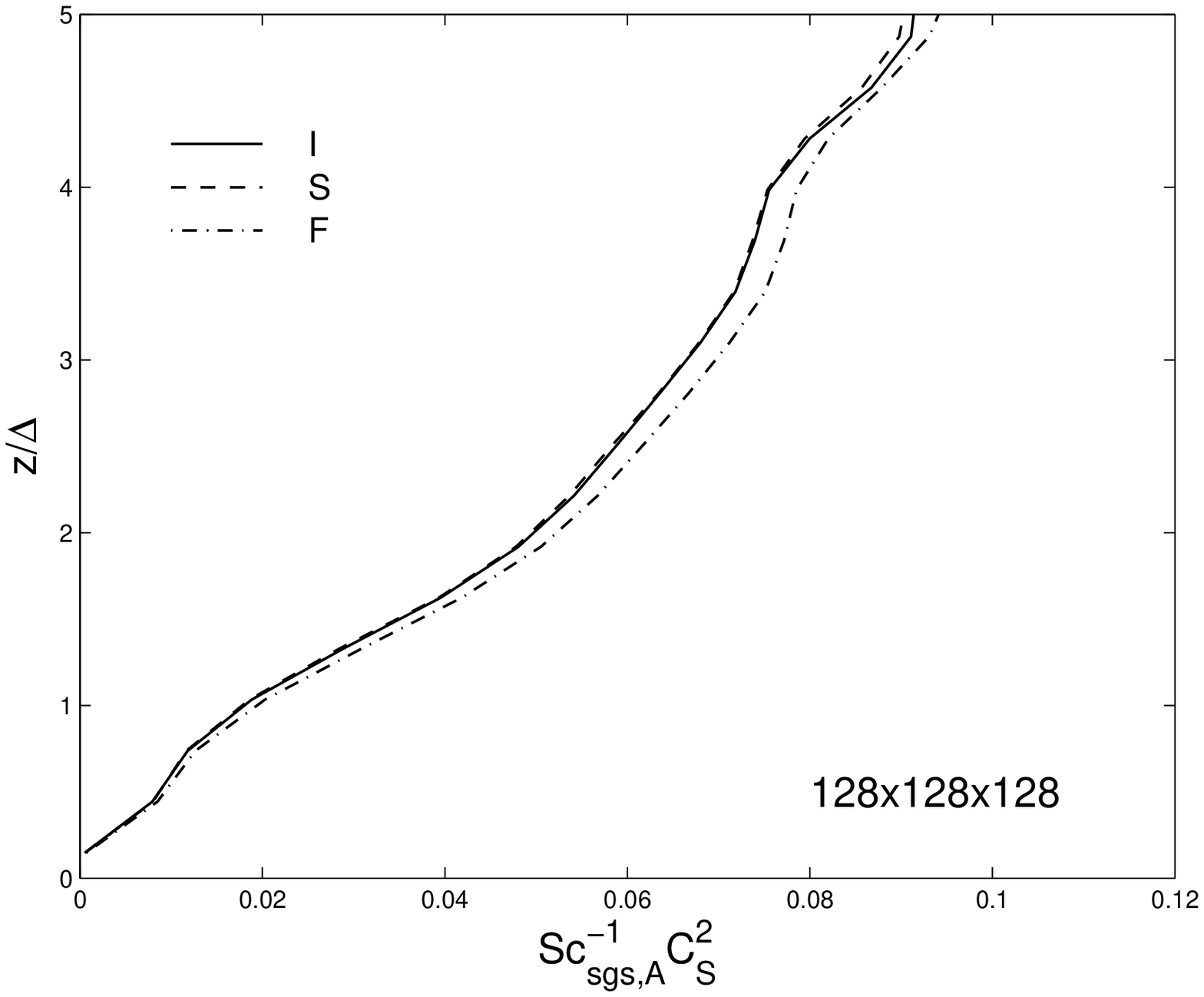}
}
\caption{Vertical profiles of $\beta_{A}$ (left column) and lumped coefficient $Sc_{sgs, A}^{-1}C_{S}^{2}$ (right
column) for the $32^{3}$ (upper row), $64^{3}$ (middle row) and $128^{3}$ (lower row) simulations.
The results have been averaged over the last hour of simulation.} \label{fig:fig4}
\end{figure}

Fig.~\ref{fig:fig4} show the vertical profiles 
of scale-dependent parameter $\beta_{A}$ and the lumped coefficient $Sc_{sgs, A}^{-1}C_{S}^{2}$
as a function of the normalized height $z/\Delta $. In order to isolate the
effect of the chemical regime from that of the resolution, the results from
each resolution are presented in different panels. One can
notice that for both $\beta _{A}$ and $Sc_{sgs,A}^{-1}C_{S}^{2}$, the results from the inert and slow chemical cases 
are the same for the highest resolution. They are
not affected by the chemistry in the $I$ and $S$ chemical cases, showing
only a scale (resolution) dependence. The little effect of the chemistry on
the subgrid fluxes in the slow chemical regime is due to the fact that the
characteristic time scale of the chemistry is much larger than the
characteristic time scale of the subgrid scale fluxes, as illustrated by the
very low values of the subgrid Damk\"{o}hler number. When decreasing
(coarsening) the resolution, however, chemistry has a relatively more
important effect on the subgrid-scale fluxes, which translates also into an
increase in the subgrid Damk\"{o}hler number. As a result, the deviation from the $I$
chemical case profiles increases when coarsening the resolution. 
Chemistry has a clear effect on all $F$ experiments with this effect increasing as the
resolution coarsens and, consequently, $Da_{sgs}$ increases. In summary, our
results show that the scale-dependent dynamic model is able to
caputure the dependence of the subgrid coefficient and scale-dependent
parameter on both resolution and chemical regime. 

\begin{figure}
\resizebox{0.75\textwidth}{!}{%
  \includegraphics{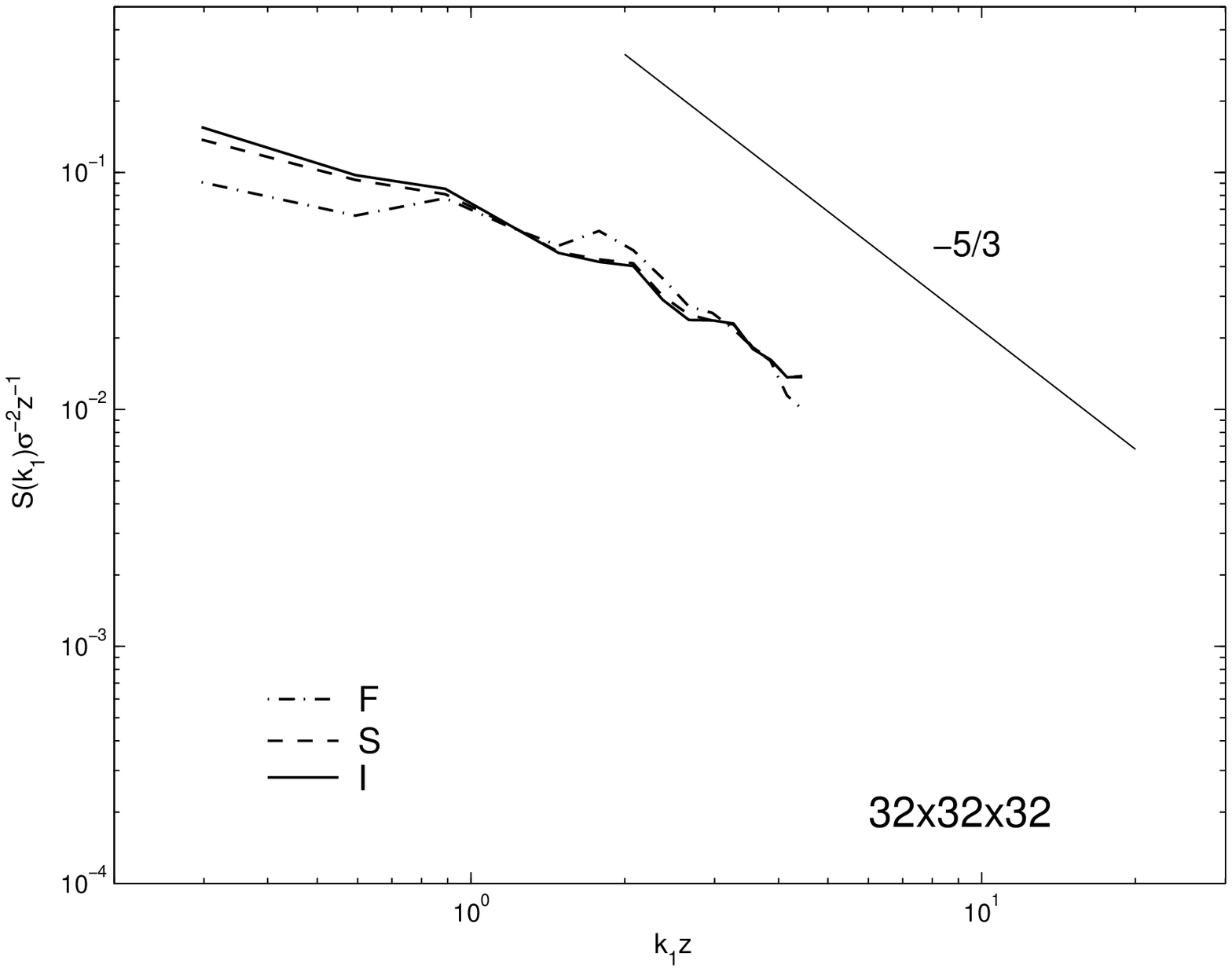}
}
\resizebox{0.75\textwidth}{!}{%
  \includegraphics{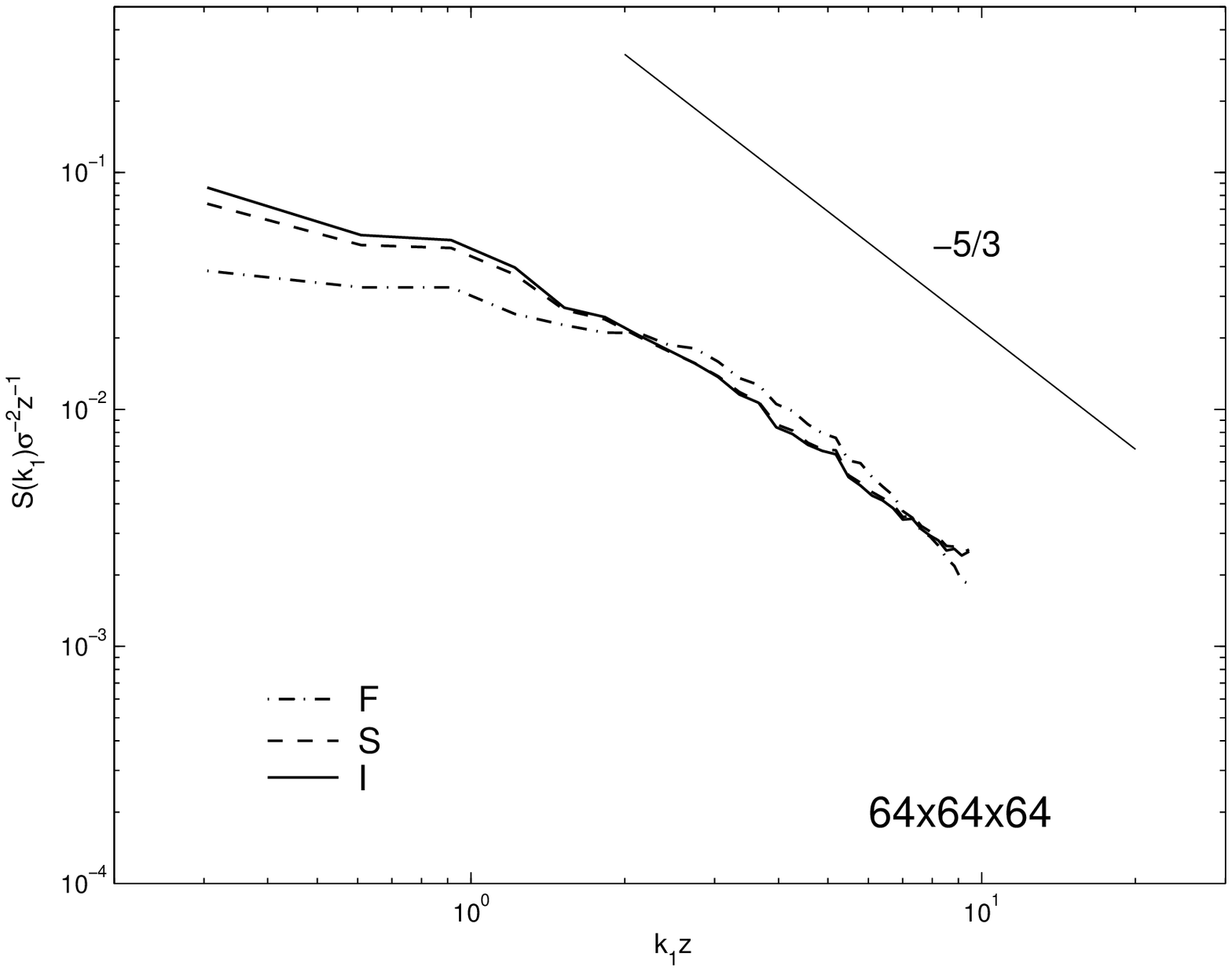}
}
\resizebox{0.75\textwidth}{!}{%
 \includegraphics{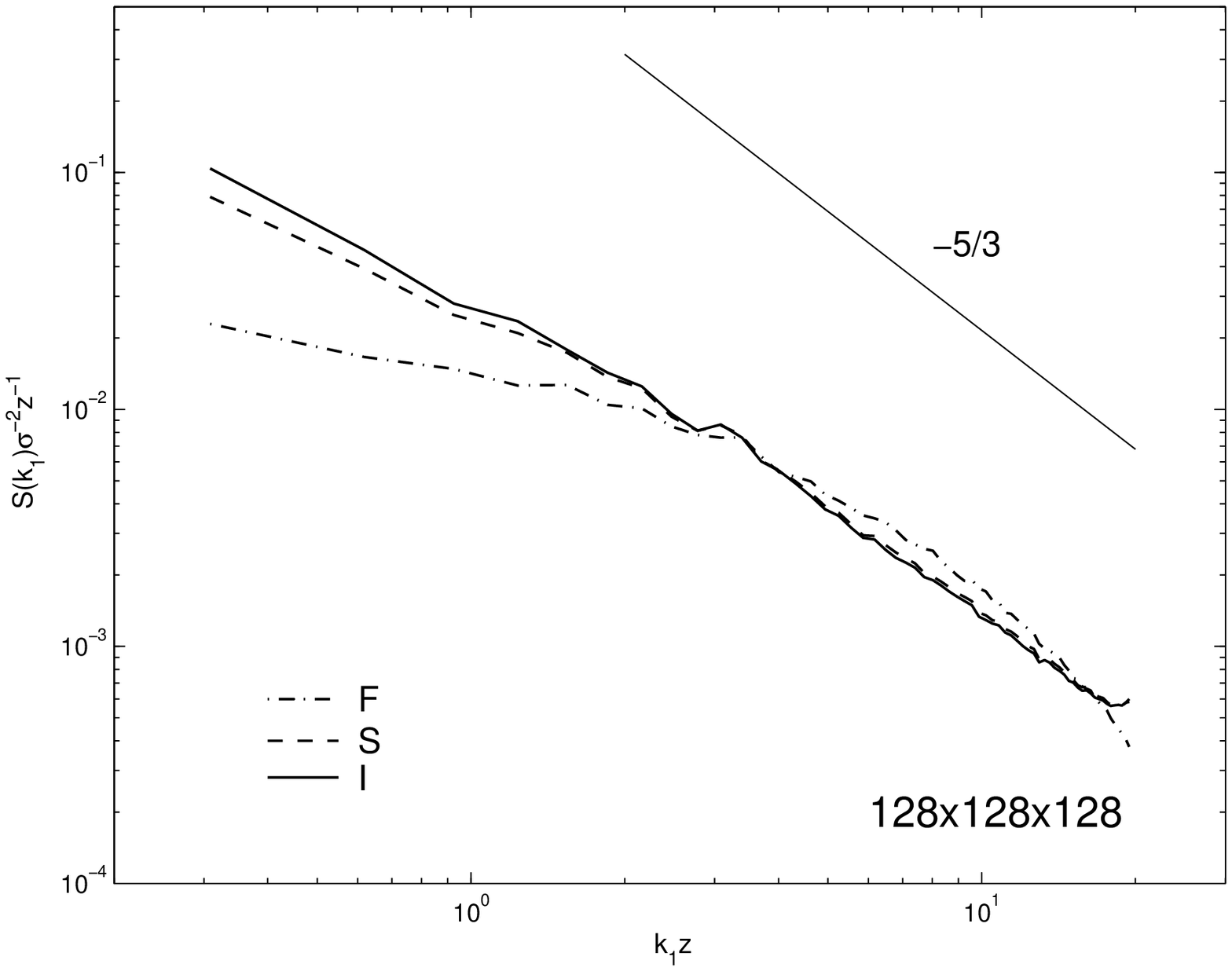}
}
\caption{Spatial variance spectra obtained for the $32^3$, $64^3$ and $128^3$ experiments (from top to bottom), normalized by $\sigma^{2}z$,
where $\sigma^{2}$ is the area below the spectral curve, calculated for a height $z = 0.3 h$. The $I$, $S$ and $F$ chemical cases
are represented by solid, dashed and dash-dotted lines, respectively. The slope
of {-5/3} is also shown.} \label{fig:fig5}
\end{figure}

The scalar variance spectra of the reactive scalar obtained from all the simulations at a height of $z=0.3h$ are shown in Fig.~\ref{fig:fig5}. 
The spectra are calculated from spatial information using one-dimensional Fourier transforms that are then
averaged in the spanwise direction and also in time. They are normalized with $\sigma^{2}z$, where $\sigma ^{2}$ is the area
below the spectral curve, and they are plotted against $k_{1}z$, where $k_{1}$ is the streamwise wavenumber. Doing so
allows us to focus on the distribution of the variance over the range of resolved scales. \cite{por04} showed
that the normalized spectra for inert scalars and heights $z\leqslant0.3h$ collapse and are proportional to $k_{1}^{-5/3}$ in
the inertial surbrange, which corresponds to scales smaller than the height (i.e., $k_{1}z\gtrsim 1$). At the largest scales
(smaller wavenumbers), our results agree with Jonker {\it et al.} \cite{jon04}, who showed that the effect of the chemistry
on the variance spectra depends on the chemical regime, characterized by the turbulent Damk\"{o}hler number
$Da_t$. For slow chemical regimes, where $Da_t << 1$, the time scale of the chemistry ($\tau_c$) is much slower than the
integral time scale of the turbulence ($\tau_t$) and, consequently, the chemical reaction has practically no effect on the
scalar variance at any of the turbulence scales. When $Da_{t}\gtrsim O(1)$, the chemistry is faster than the characteristic
turning time of the largest eddies, which has an impact (reduction) on the level of fluctuations of the reacting
scalars at those scales. As a consequence, in Fig. ~\ref{fig:fig5} the spectra corresponding to the fast chemical regime
show substantially smaller variance at the largest scales (small $k_1$) for all the resolutions. In order to understand
if that effect will extend to the subgrid and/or smallest resolved scales, one needs to consider, in addition, the
value of the subgrid Damk\"{o}hler number $Da_{sgs}$. If $Da_{sgs} << 1$ then the subgrid and smallest resolved scales
are not affected and, consequently, the scale-dependent dynamic model parameters are not expected to depart
from the inert scalar values, as shown in Fig.~\ref{fig:fig4}. On the other hand, if $Da_{sgs} \gtrsim  O(1)$, then the effect
of the chemistry extends also to the subgrid and/or smallest resolved scales, which become less isotropic. It
is important to note that the tuning-free dynamic model adjusts to the less isotropic behavior of those scales
through a reduction of the model parameters as shown in Fig.~\ref{fig:fig4}.

\section{Summary}
\label{sum}

A scale-dependent dynamic model for the subgrid-scale fluxes of reacting
scalars is presented and used in large-eddy simulations of atmospheric
boundary layer reacting flows. This model uses a dynamic procedure to
calculate the lumped coefficient ($Sc_{sgs,\psi }^{-1}C_{S}^{2}$) in the
eddy-diffusion subgrid-scale model as a function of the dynamics of the
resolved scales. Consequently, the proposed dynamic model does not require
any parameter specification or a priori tuning.

The scale-dependent dynamic model is implemented in simulations of a neutral
ABL with a constant and uniform surface flux of a first-order decaying
scalar. A set of spatial resolutions and reaction rates are used in order to
study the ability of the dynamic model to adjust the model coefficient to
changes in scale (resolution) and/or chemical regime. The model is able to
account for the scale dependence of the model coefficient in a self
consistent way. Scale dependence increases with decreasing
height-to-scale ratio due to the increased flow anisotropy at the filter
and/or test filter scales. In addition, we found that when the time scale
associated with the subgrid turbulence is on the same order of magnitude
that the time scale associated with the chemistry, i.e. $Da_{sgs}\sim O(1)$,
an additional dependence towards the chemical reactivity can be expected. By
calculating explicitly the lumped model coefficients for the reacting
scalars, there is no need for any assumption on the combined effects of
resolution and reaction rate on the efficiency of the subgrid scale
turbulence to transport scalars. We showed that the model is able to capture
the chemical regime dependence of the reactant and, as a result, simulations
yield resolution-independent total turbulent reactant fluxes.
%
%

\begin{thebibliography}{}
%
%


\bibitem[1]{kle97}
Kley, D.,:1997, Tropospheric chemistry and transport,
{\it Science} {\bf 276}, 1043--1047.

\bibitem[2]{sch89}
Schumann, U.: 1989, Large-eddy simulation of turbulent diffusion
with chemical reactions in the convective boundary layer,
{\it Atmos. Env.} {\bf 23}, 1713--1729.

\bibitem[3]{gao93}
Gao, W., Wesely, M.~L., Doskey, P.~V.: 1993, 
Numerical modeling of the turbulent diffusion and chemistry of NO$_{x}$, O$_{3}$, isoprene and other reactive trace gases in and above a forest canopy,
{\it J. Geophys. Res.} {\bf 98}, 18339--18353.

\bibitem[4]{syk94}
Sykes, R.~I., Parker, S.~F., Henn, D.~S., Lewellen, W.~S.: 1994,
Turbulent mixing with chemical reactions in the planetary boundary layer, 
{\it J. Applied Meteorol.} {\bf 33}, 825--834.

\bibitem[5]{gao94}
Gao, W., Wesely, M.~L.: 1994, 
Numerical modelling of the turbulent fluxes of chemically reactive trace gases in the atmospheric boundary layer, 
{\it J. Applied Meteorol.} {\bf 33}, 835--847.

\bibitem[6]{ver97}
Verver, G. H.~L., van Dop, H., Holtslag, A.~A.~M.: 1997,
Turbulent mixing of reactive gases in the convective boundary layer, 
{\it Bound.-Layer Meteorol.} {\bf 85}, 197--222.

\bibitem[7]{mol98}
Molemaker, M.~J., {Vil\`{a}-Guerau de Arellano}, J.: 1998,
Turbulent control of chemical reactions in the convective boundary layer, 
{\it J. Atmos. Sci.} {\bf 55}, 568--579.

\bibitem[8]{pet99b}
Petersen, A.~C., Beets, C., van Dop, H., Duynkerke, P.~G.: 1999,
Mass-flux schemes for transport of non-reactive and reactive scalars in the convective boundary layer, 
{\it J. Atmos. Sci.} {\bf 56}, 37--56.

\bibitem[9]{pet99c}
Petersen, A.~C., Holtslag, A.~A.~M.: 1999, 
A first-order closure for covariances and fluxes of reactive species in the convective boundary layer, 
{\it J. Appl. Meteorol.} {\bf 38}, 1758--1776.

\bibitem[10]{pet00}
Petersen, A.~C.: 2000, 
The impact of chemistry on flux estimates in the convective boundary layer, 
{\it J. Atmos. Sci.} {\bf 57}, 3398--3405.

\bibitem[11]{kro00}
Krol, M.~C., Molemaker, M.~J., {Vil\`{a}-Guerau de Arellano}, J.: 2000, 
Effects of turbulence and heterogeneous emissions on photochemically active species in the convective boundary layer,
{\it J. Geophys. Res.} {\bf 105}, 6871--6884.

\bibitem[12]{ver00}
Verver, G.~H.~L., van Dop, H., Holtslag, A.~A.~M.: 2000,
Turbulent mixing and the chemical breakdown of isoprene in the atmospheric boundary layer, 
{\it J. Geophys. Res.} {\bf 105}, 3983--4002.

\bibitem[13]{vil00}
{Vil\`{a}-Guerau de Arellano}, J., Cuijpers, J.~W.~M.: 2000, 
The chemistry of a dry cloud: the effects of radiation and turbulence,
{\it J. Atmos. Sci.} {\bf 57}, 1573--1584.

\bibitem[14]{pat01}
Patton, E.~G., Davis, K.~J., Barth, M.~C., Sullivan, P.~P.: 2001,
Decaying scalars emitted by a forest canopy: a numerical study,
{\it Bound.-Layer Meteorol.} {\bf 100}, 91--129.

\bibitem[15]{vin03}
Vinuesa, J.-F., {Vil\`{a}-Guerau de Arellano}, J.: 2003, Fluxes
and (co-)variances of reacting scalars in the convective boundary
layer, {\it Tellus} {\bf 55B}, 935--949.

\bibitem[16]{jon04}
Jonker, H.~J.~J., {Vil\`{a}-Guerau de Arellano}, J., Duynkerke, P.~G.: 2004,
Characteristic length scales of reactive species in a convective boundary layer,
{\it J. Atmos. Sci.} {\bf 61}, 41--56.

\bibitem[17]{vin04}
Vinuesa, J.-F., {Vil\`{a}-Guerau de Arellano}, J.: 2005, Introducing effective 
reaction rates to account for the inefficient mixing of the convective boundary layer, 
{\it Atmos. Env.} {\bf 39}, 445--461.

\bibitem[18]{moe84}
Moeng, C.~H.: 1984, 
A large-eddy simulation model for the study of planetary boundary-layer turbulence,
{\it J. Atmos. Sci.} {\bf 41}, 2052--2062.

\bibitem[19]{vin05}
Vinuesa, J.-F., {Port\'{e}-Agel}, F.: 2005, 
A dynamic similarity subgrid model for chemical transformations in Large Eddy Simulation of the atmospheric boundary layer,
{\it Geophys. Res. Let.} {\bf 32}, L03814.

\bibitem[20]{por04}
Port\'{e}-Agel, F.: 2004,
A scale dependent dynamic model for scalar transport in LES of the atmospheric boundary layer, 
{\it Bound.-Layer Meteorol.} {\bf 112}, 81--105.

\bibitem[21]{mee00}
Meeder, J.~P., Nieuwstadt, F.~T.~M.: 2000,
Large-eddy simulation of the turbulent dispersion of a reactive plume from
a point source into a neutral atmospheric boundary layer,
{\it Atmos. Env.} {\bf 34}, 3563--3573.

\bibitem[22]{lil67}
Lilly, D.~K.: 1967, 
The representation of small-scale turbulence in numerical simulation experiments,
{\it Proc. IBM Scientific Computing Symposium on Environmental Sciences}, IBM form no. 320-1951, White Plains, New-York, 195--209.

\bibitem[23]{mas90}
Masson, P. J., Derbyshire, S. H.: 1990, 
Large-eddy simulation of the stably-stratified atmospheric boundary layer,
{\it Bound.-Layer Meteorol.} {\bf 53}, 117--162.

\bibitem[24]{ger92} 
Germano, M.: 1992, 
Turbulence: the filtering approach,
{\it J. Fluid Mech.} {\bf 238}, 325--336.

\bibitem[25]{lil92}
Lilly, D.~K.: 1992, 
A proposed modification of the Germano subgrid-scale closure method,
{\it Phys. Fluids A} {\bf 4}, 633--635.

\bibitem[26]{vil03a}
{Vil\`{a}-Guerau de Arellano}, J.: 2003, Bridging the gap between
atmospheric physics and chemistry in studies of small-scale
turbulence, {\it Bulletin of the American Meteorological Society} {\bf 84}, 51--56.

\bibitem[27]{pop00}
Pope, S.~B.: 2000, Turbulent flows,
Cambridge University Press.

\bibitem[28]{men00}
Meneveau, C., Katz, J.: 2000, Scale-invariance and turbulence models for 
large-eddy simulation,
{\it Rev. Fluid Mech.} {\bf 32}, 1--32.

\bibitem[29]{alb99}
Albertson, J.~D., Parlange, M.~B.: 1999,
Natural integration of scalar fluxes from complex terrain,
{\it Adv. Wat. Res.} {\bf 23}, 239--252.

\bibitem[30]{por00}
Port{\'e}-Agel, F., Meneveau, C., Parlange, M.~B.: 2000, 
A scale-dependent dynamic model for large-eddy simulation: application to a neutral atmospheric boundary layer, 
{\it J. Fluid Mech.} {\bf 415}, 261--284.

\end{thebibliography}
%

\acknowledgement{
The authors gratefully acknowledge the financial support by
NASA (NAG5-11801), the National Science Foundation (EAR-0094200
and EAR-0120914 as part of the National Center for Earth-Surface Dynamics). Computational 
resources were provided by the Supercomputing Institute for Digital Simulation and Advanced
Computation (MSI). J.-F. Vinuesa received partial support from a MSI research fellowship.
}


\end{document}